\pdfoutput=1
\let\EXTEND=y

\documentclass[sigconf]{acmart}
\usepackage{tikz}
\usepackage{amsmath}
\usepackage{textcomp}
\usepackage{verbatim}
\usepackage{xspace}
\usepackage{multirow}
\usepackage{graphicx}
\usepackage{subfig}
\usepackage{makecell}
\usepackage{balance}
\usepackage{url}

\AtEndPreamble{\hypersetup{colorlinks,
linkcolor=ACMPurple,
citecolor=ACMPurple,
urlcolor=ACMDarkBlue,
filecolor=ACMDarkBlue}}

\ifx\MARKED y
\newcommand{\mr}[1]{\textcolor{blue}{#1}}
\newenvironment{mrlong}{\color{blue}}{}
\else
\newcommand{\mr}[1]{#1}
\newenvironment{mrlong}{}{}
\fi

\newcommand{\Name}{InviCloak\xspace}
\newcommand{\name}{InviCloak\xspace}

\newcommand{\eg}{{\it e.g.}\xspace}
\newcommand{\ie}{{\it i.e.}\xspace}

\newcommand{\None}{\makecell[c]{-}}
\newcommand{\tabhdr}[1]{\makecell[c]{\textbf{#1}}}
\newcommand{\tabhdrTwo}[2]{\makecell[c]{\textbf{#1} \\ \textbf{#2}}}

\usepackage{enumitem}

\AtBeginDocument{%
  }

\copyrightyear{2022}
\acmYear{2022}
\ifx\EXTEND y
    \setcopyright{none}
\else
    \setcopyright{acmcopyright}
\fi
\acmConference[CCS '22]{Proceedings of the 2022 ACM SIGSAC Conference on Computer and Communications Security}{November 7--11, 2022}{Los Angeles, CA, USA}
\acmBooktitle{Proceedings of the 2022 ACM SIGSAC Conference on Computer and Communications Security (CCS '22), November 7--11, 2022, Los Angeles, CA, USA}
\acmPrice{15.00}
\acmDOI{10.1145/3548606.3559336}
\acmISBN{978-1-4503-9450-5/22/11}

\ifx\EXTEND y
    \usepackage{tabularx}
\else
\fi

\acmSubmissionID{10}

\begin{document}
\epstopdfsetup{suffix=,}

\title{\Name: An End-to-End Approach to Privacy and Performance in Web Content Distribution}

\settopmatter{authorsperrow=4}
\author{Shihan Lin}
\affiliation{%
    \institution{Duke University}
    \country{}
}

\author{Rui Xin}
\affiliation{%
    \institution{Duke University}
    \country{}
}

\author{Aayush Goel}
\affiliation{%
    \institution{Cupertino High School}
    \country{}
}

\author{Xiaowei Yang}
\affiliation{%
    \institution{Duke University}
    \country{}
}


\begin{abstract}


    %

    In today's web ecosystem, a website that uses a Content
    Delivery Network (CDN) shares its Transport Layer Security
    (TLS) private key or session key with the CDN. In this paper, we
    present the design and implementation of \name, a system that
    protects the confidentiality and integrity of a user and a
    website's private communications without changing TLS or upgrading a CDN.
    \Name builds a lightweight but secure and practical key
    distribution mechanism using the existing DNS infrastructure
    to distribute a new public key associated with a website's domain name.
    A web client and a website can use the new key pair to build an encryption
    channel inside TLS. \Name accommodates the current web ecosystem.
    A website can deploy \name unilaterally
    without a client's involvement to prevent a passive attacker
    inside a CDN from eavesdropping on their communications.  If a
    client also installs \name's browser extension, the client and
    the website can achieve end-to-end confidential and untampered
    communications in the presence of an active attacker inside a
    CDN. \mr{
        Our evaluation shows that \name increases the median page load
        times (PLTs) of realistic web pages from 2.0s to 2.1s,
        which is smaller than the median PLTs (2.8s) of a state-of-the-art TEE-based solution.
    }





\end{abstract}




\maketitle

\section{Introduction}\label{Introduction}

Content Delivery Networks (CDNs) play an important role in the web
ecosystem. They not only speed up web content distribution but also
protect a website from a wide range of attacks. For example, CDNs
such as Akamai~\cite{nygren2010akamai}, Cloudflare~\cite{Cloudflare},
and CloudFront~\cite{CloudFront} offer Distributed Denial of Service
(DDoS) attack mitigation and malicious content scrubbing
services~\cite{gillman2015protecting,NGINXProxyModule}.

Unfortunately, as more and more websites migrate to 100\%
HTTPS \cite{HTTPSStats}, a third-party CDN introduces undesirable
security ramifications.  It is a common practice for a website to share
its Transport Layer Security (TLS) certificate's private key with a
CDN to fully take advantage of a CDN's performance and security
benefits. A measurement conducted by Cangialosi et al. in 2016 shows
that 76.5\% of organizations share their private keys with a
third-party hosting provider, and for popular websites, they mainly
share their keys with CDN providers\cite{cangialosi2016measurement}.

This key-sharing practice breaks the end-to-end security guarantees
offered by TLS/HTTPS, posing potential security risks. A CDN, as an
organization, may suffer from an insider attack or exploitable security
vulnerabilities. Since a CDN typically serves many websites, a
compromised CDN may leak the private credentials of many web services,
becoming a central point of failure. As an example, web cache
deception attacks~\cite{Security20-WCD,WCD} exploit a CDN's configuration
vulnerability. Researchers have shown that an attacker can deceive a
CDN into caching and exposing Personally Identifiable Information
(PII) such as names and phone numbers~\cite{Security20-WCD}.


The security risk of this current practice has prompted multiple
solutions. Each has made trade-offs among adoptability, security, and
performance. Cloudflare's Keyless SSL~\cite{CloudflareKeylessSSL} and
certificate delegation~\cite{liang2014https} do not expose a website's
TLS private key to a CDN, but still allow a CDN to possess the TLS
session key -- this allows an attacker inside a CDN to continue to
observe and modify the content inside a TLS session. Other solutions
such as mcTLS and
maTLS~\cite{SIGCOMM15-mcTLS,SP18-formal,NDSS19-maTLS} modify
the TLS protocol to include middleboxes in a TLS handshake.  Although
technically sound, these solutions require coordinated efforts from clients, middleboxes, and websites to upgrade their TLS implementations.


Alternatively, a website could obtain two TLS certificates for two
different domains (\eg, \url{site.com} and \url{site-cdn.com}): one
for privacy-sensitive content hosted by itself and the other for
content hosted by a CDN, similar to~\cite{NDSS16-CDNonDemand}. The
website shares the CDN-related TLS certificate's private key with its
CDN and keeps the other one private. We refer to this proposal as
\emph{the two-domain solution}. A main drawback of the two-domain
solution is that it does not protect against active attacks when a
website uses a CDN to distribute its base HTML file, which is the
first file a user downloads when she visits a web page. For
performance reasons, a website desires to distribute its base HTML
file via a CDN~\cite{AkamaiHTMLDelivery}(\S~\ref{Motivation}), but an
active attacker inside the CDN could modify this file and hijack the
subsequent private TLS sessions.  Moreover, the two-domain
solution prevents a CDN from caching any private content, even in encrypted form, as a website
will send it via a separate TLS session. This  design reduces the performance benefit of a CDN.

In a different direction, researchers have proposed to use Trusted
Execution Environments
(TEEs)~\cite{anati2013innovative,costan2016intel} to prevent
distrusted CDN code from accessing a shared TLS private
key~\cite{wei2017styx,ahmed2018harpocrates} or a TLS session
key~\cite{Security20-Phoneix}.  These solutions offer desirable
security guarantees without any deployment effort on the user side.
However, they face both deployment and performance challenges for
CDNs. Because of costly system calls inside an enclave,
the current TEE hardware may slow down a CDN edge
server's throughput by two to four times~\cite{Security20-Phoneix}.  In
addition, these solutions require CDNs to upgrade the hardware of
their infrastructure. The financial cost of upgrading a CDN's
infrastructure to support a TEE-based solution can exceed more than
100 million dollars per our analysis (\S~\ref{DeploymentCost}).
Furthermore, the future of current TEE-based solutions is unclear,
since Intel has announced that the Intel TEE,
called Software Guard Extensions (SGX)~\cite{costan2016intel} is deprecated
in the 12th generation of Intel CPUs~\cite{DeprecatedTechnologies}.

Each of the existing proposals has its own security, performance, and
deployment cost trade-offs. In this work, we aim to explore a solution
with different cost and benefit trade-offs for the market to choose
from. The solution, \name, takes an end-to-end approach.  It
accommodates the current key-sharing practice and separates content
serving authorization from confidentiality.  A website uses the shared
TLS key to authorize a CDN to serve its non-privacy-sensitive
content. It then uses a new pair of private/public keys that it does
not share with a CDN to protect privacy-sensitive content. \Name
protects against active attacks and does not increase the traffic sent
to an origin server.

%


A main design challenge \name faces is how to balance the
security benefit it brings with its deployment and performance
cost. To address this challenge, we design \name to  use the existing
DNSSEC~\cite{RFC4033-DNSSEC} and DNS-over-HTTPS~\cite{RFC8484-DoH}
(DoH) infrastructure to distribute a website's new public key, thereby
obviating the need for the website to obtain a new TLS certificate
for a new domain name.
For ease of deployment, \name embeds an end-to-end
encryption tunnel inside the existing TLS sessions between a web
client and a website's origin server to transmit private
data, such as a user's login password. This design obviates the need
for modifying TLS, a CDN, a web server, or web resources.

As a result, \name has several deployment advantages over existing
proposals. It does not change the underlying TLS protocol and is
completely transparent to a CDN. CDNs need not upgrade their
infrastructure. Furthermore, it does not modify existing web
server implementations, and web developers need not change existing
web resources or manage new domain names and certificates. A website
can unilaterally deploy \name as a JavaScript library to defeat a
passive eavesdropper without any user-side operations. If a user
installs \name's browser extension, she can detect and prevent an
active attacker from tampering or eavesdropping on her private
communications with a website.



As performance is critical to web applications, the encryption-in-encryption design of
\name is easy to be dismissed due to its overhead.
We implemented a prototype, micro-benchmarked \name's operations,
and measured how it affects the page load times (PLTs) of web pages.
\mr{On our testbed, \name introduces less than 100~ms delay to median PLTs. This overhead
    is about 700~ms lower than a state-of-the-art TEE-based
    solution~\cite{Security20-Phoneix}.} If the overhead becomes a concern,
we can modify browser implementations to eliminate the inner-layer
encryption at the cost of an increased deployment hurdle.


%


This work makes the following key contributions:
\begin{itemize}
    \item The design of \name, which protects users' private data
          from a compromised CDN while keeping the CDN functioning as a
          DDoS shield for a website.
    \item A prototype implementation that is immediately deployable
          within the current web ecosystem~\footnote{The source code is accessible at \url{https://github.com/SHiftLin/CCS2022-InviCloak}}.
          Our evaluation shows that it introduces acceptable overhead to web content distribution.
    \item We analyze the deployment efforts of \name and compare it
          with related work. We show that \name's deployment
          requires no modifications of a CDN, TLS, OSes, or a web server.
          Neither does it require a new domain name
          nor a new TLS certificate.
\end{itemize}

\textbf{Ethical concern:} This work does not raise any ethical concerns.

\section{Design Rationale}\label{ProblemDefinition}


\subsection{Motivation}\label{Motivation}
We have conducted a measurement study on Alexa top-100 websites~\cite{AlexaTopSites}
to understand how websites that use third-party CDNs protect
their privacy-sensitive data such as user login passwords.

We refer to methodologies in existing research~\cite{cangialosi2016measurement,BSThesisPrinceston17-CDNPrivacy,huang8measuring,krishnamurthy2001use} to discover the CDN usage of a website and determine the organization to which a website belongs.
If a website's organization is not the same as the CDN provider, we
conclude that the website uses a third-party CDN.
The result shows that 67 of top-100 websites employ third-party CDNs, and 54
of them use the CDNs to deliver the homepages' base HTML files. For
all these 54 websites, an active attacker
inside a CDN could modify those pages even if they employ the two-domain solution as described in \S~\ref{Introduction}.


We further examined the login procedure of these websites to
investigate their strategies for sensitive data transmission. Of the
67 CDN-enabled websites, 7 of them do not use password logins, and
25 of them expose their login servers' IP addresses in the login
procedure, while 35 of them send users' passwords through CDNs.
This result suggests that not all websites trust the CDNs they use,
as indicated by the 25 websites which bypass CDNs for user
logins. However, exposing the IP address of a website's login server
makes the website vulnerable to DDoS attacks.
For those websites that expose their users' passwords to
CDNs, they risk leaking users' sensitive and private data
to the CDNs.

The above observations motivate us to design a solution that is both
conceptually simple and secure.




\subsection{CDN Service Model}
For clarity, we describe a service model between a website and its CDN
service provider. We use this model to design \name. Specifically, we
categorize the content served by a website into \emph{CDN-visible}
content and \emph{private} content. We regard private content as
the content that belongs to a registered user of a website and should only
be accessible to an authenticated user. For example, a user must log
in to check her bank account balance. If some content is cached by a
CDN or is not private to a user, we consider it CDN-visible.  For
example, static content cached by CDN servers or behind a
paywall,
such as videos available from a subscription service, is
CDN-visible.


Due to privacy concerns, a website does not share its
private databases with a CDN. It uses a CDN to cache and serve its
CDN-visible content, but a user will send/retrieve private content
directly to/from a website's origin server~\cite{CloudflareCacheExplained}.
Furthermore, we assume it is desirable for a website to cache some private content
such as private user photos on CDNs for performance acceleration.
Thus, \Name's design also supports such use cases while keeping the
private content secret to CDNs.  We note that \name
\emph{does not} change the existing CDN service model so it \emph{will
    not} increase the traffic volume sent to a website's origin server.
A website that does not serve any private content is outside the
scope of this work and does not need to deploy \name.

In this paper, we refer to a server hosted by a website as an
``\emph{origin server}'', which is the initial source of all content
of the website. The term ``\emph{user}'' always denotes the user of a
website instead of a CDN. When we use the term ``\emph{client}'', it
refers to the endpoint (a browser or a computer) that a user uses.
Finally, an ``\emph{attacker}'' refers to any malicious entity.



\subsection{Threat Model}\label{ThreatModel}

We assume that not all CDN customers completely trust their
CDNs. These customers would like to benefit from CDN's caching and
DDoS protection services without exposing private content to
CDNs. We assume a compromised CDN may launch two types of attacks with
different risk factors.



\textbf{Passive Attacks:} A CDN may have a software or
configuration vulnerability that results in unintended information
leak, such as the web cache deception attack~\cite{Security20-WCD,WCD}.
Or a curious eavesdropper inside the CDN may log the
plaintext data after a CDN's edge server decrypts the TLS session
data it receives.  We model this type of vulnerability as passive attacks
that eavesdrop on private communications between a web
client and a website's origin server.

%

\textbf{Active Attacks:} A CDN may have a compromised insider
that gains access to its customer websites' TLS private keys, or there
is a software bug inside the CDN that allows an attacker to inject
malicious code. We model this threat as active attacks that can
eavesdrop, tamper, or leak any message it receives.



\begin{figure*}[th]
    \begin{center}
        \includegraphics[width=0.75\textwidth]{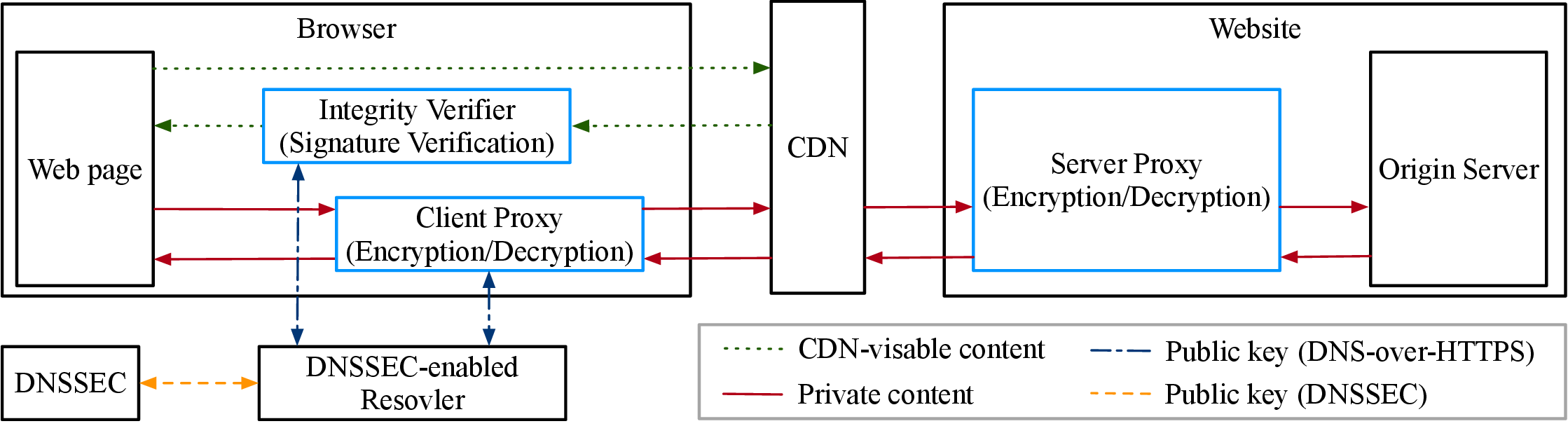}
        \caption{\textbf{\small This figure shows the high-level
                design of \Name in the case of active attackers.
                It has three components: a client proxy,
                a server proxy, and an integrity verifier.
                The client proxy and server proxy encrypt and decrypt the
                communication between a client and a website's origin
                server, while the integrity verifier validates the
                integrity of HTML and JavaScript files returned by a CDN.
                \mr{The client proxy is a Service Worker,
                    which is different from regular JavaScript code (\S~\ref{Design:ClientProxy})}. In the case of passive attackers, the integrity verifier is unnecessary.
            }}
        \label{fig:architecture}
    \end{center}
\end{figure*}

\subsection{Trust Assumptions}\label{Assumptions}

We make the following trust assumptions.

\textbf{No colluding attackers between DNS and CDNs:}
We assume that an adversary is not powerful enough to
penetrate both a website's CDN provider and its DNS provider. In
other words, even if there exists an attacker in a website's DNS
provider and one in its CDN provider, we assume the two adversaries
are independent of each other and cannot collude with each
other.

\Name cannot defeat a powerful attacker that compromises both a
website's DNS and CDN providers since its design uses DNS to
distribute a new public key. Thus, if a web service is concerned with
colluding DNS and CDN attackers, it can choose to separate its CDN
provider from its DNS provider. In practice, all major CDNs we
survey, including Akamai, CloudFront, Google Cloud, Azure,
and Fastly, allow users to use separate DNS providers.
Cloudflare, by default, requires its
customers to use their DNS service to host the customers' domain
names, but it also allows customers to switch to other DNS
providers~\cite{CloudflareCNAME}.

\begin{mrlong}
    We conducted a measurement study on Alexa top-10K websites to validate this assumption.
    We first used the existing methods~\cite{cangialosi2016measurement,BSThesisPrinceston17-CDNPrivacy,huang8measuring,krishnamurthy2001use}
    to discover the third-party CDN usage of the websites.
    Moreover, we used \texttt{dig}~\cite{LinuxDIG} to obtain the nameservers of each CDN-enabled domain in a website,
    and we obtained the domain registrars from the WHOIS service~\cite{RFC3912-WHOIS}. 
    A website may own multiple domains.
    If the nameserver of a domain does not belong to the website's CDN provider,
    we regard that the website separates the DNS provider from the CDN provider.
    Besides, if the domain registrar is not the CDN provider,
    we regard that the website separates the domain registrar from the CDN provider.

    Among the top-10K websites, 4867 of them use a third-party CDN provider.
    We find that 2765 of the 4867 (57\%) websites already separated their DNS providers from their CDN providers.
    Moreover, among those 2102 (43\%) websites that do not have the separation, 1668
    of them separate their domain registrars from CDNs. These 1668 websites
    can separate their DNS providers from their CDNs by transferring their
    DNS providers to their domain registrars without financial cost.
\end{mrlong}

\textbf{Bootstrapping security:} We assume users or websites can
obtain their OSes, browsers, and \name-related modules or extensions securely
without involving a compromised CDN.

\textbf{Trusted Computing Base (TCB):} \mr{We consider the implementations of the browsers, websites, OSes,  and hardware of the clients and origin servers as our TCB. Specifically, we trust the implementation of a website that would deploy \name, including its web pages (HTML, JavaScript, CSS) and all other system components it uses such as databases and the service backend.
    Admittedly, an attacker may exploit the vulnerabilities in a
    website or a client's browser to launch attacks such as SQL
    injection~\cite{halfond2006classification} and cross-site
    scripting (XSS)~\cite{weinberger2011systematic}. We consider
    preventing such attacks outside the scope of this work.}

%
%

\textbf{Hardness of cryptography:} We assume attackers
cannot overcome the hardness of a cryptographic algorithm. For
example, they cannot decrypt ciphertext without the cryptographic key,
and they cannot falsify a digital signature without the private key.

\subsection{Design Goals}

Our design goals are multi-fold, including privacy and
confidentiality, usability, low deployment cost, and performance.
These goals set our work apart from related work.

\textbf{Privacy and confidentiality:} Our foremost goal is
privacy and confidentiality.
We aim to protect private content transmitted between users and websites from leaking to third-party CDNs.
Our design should resist both active and passive attacks from a
compromised CDN.

\textbf{Usability:} An important design goal of \name is to
comply with the current usability model of the web. We require that a
website need not obtain a new domain name nor negotiate a new type of service with its CDN provider. Besides, a website should not change the domain names
displayed to its users, otherwise it may cause brand name confusion and phishing vulnerabilities.


%

\textbf{Low deployment cost:} We hypothesize that if a security
feature is financially costly and requires coordinated upgrades from
multiple stakeholders, then it is difficult to deploy that feature in
practice. Thus, we aim for low financial cost of deploying
\name. Besides financial cost, we target at minimum operation cost
for web developers and users. For example, web developers need not modify
their application logic or existing web resources.
A website need not revoke the certificate 
it shares with a CDN nor apply to a new one,
since the revocation and application of
business certificates could be time-consuming~\cite{liang2014https}.
We also aim for an end-to-end design that a website and users can
deploy without CDN support.

\textbf{Performance:} Our solution should retain a CDN's
performance benefits such as low latency and high throughput, as well
as its security benefits such as DDoS mitigation.
%


%


\section{Design}\label{Design}


\subsection{Architecture}

The main design challenge we face is to achieve \name's security goals
without sacrificing usability, deployability, or performance. We make a few design
decisions to address this challenge. First, we use the DNS-based
Authentication of Named Entities (DANE)~\cite{RFC7671-DANE} to
distribute a website's new public key to ease key management.  Second,
we design \name's main client component to be a JavaScript library so
a website can distribute it via its landing page without having the
users install it. Last, we use a reverse server proxy to serve
\name' server-side functions, so web servers need not be modified.

Figure~\ref{fig:architecture} depicts the overall architecture of
\name, its key components, and their locations. \Name introduces three
new components: a client proxy running in a web browser, an integrity
verifier introduced as a browser extension, and a server proxy running
as a reverse proxy in front of a website's origin server.  The
integrity verifier works for all \name-enabled websites, while the
client proxy is a JavaScript library that includes per website-specific
configurations. A website embeds the client proxy in its landing page,
and the integrity verifier will be distributed via a
browser-vendor-approved mechanism such as the extension market.
We note that in the case of passive attackers, the integrity verifier is unnecessary.
\Name uses the DNS infrastructure
to distribute a public key of a website for establishing an encryption
channel between a browser and the website's origin server. Its
design is completely transparent to a CDN.


We use an example to illustrate how it works at a high
level. In this example, a user named Alice visits her bank, \url{bank.com}.
We assume she is a security-conscious user and has installed \name's
integrity verifier in her web browser. 
First, Alice's browser fetches the landing page \url{bank.com/login.html} from
\url{bank.com}'s CDN provider. This page will automatically download
\url{bank.com}'s client proxy code (JavaScript). This download will trigger the
integrity verifier to send a DNS-over-HTTPS query to a DNSSEC-enabled resolver 
to obtain the public key of \url{bank.com}. 
The public key will be cached in the extension's storage for reuse. The integrity verifier validates the client proxy and installs it in the browser.

After Alice fills in her username and password for her bank account,
she clicks the submit button. This action triggers the client proxy to
encrypt the request, as the submission URL is listed as one that
contains private information and needs encryption in the client
proxy's configuration file. Since the client proxy and the
integrity verifier run as separate processes, the client proxy launches
its own DNS query to obtain \url{bank.com}'s public key. It caches
the public key in the browser's cache storage for reuse.
The client proxy then invokes
a key exchange process with the server proxy and generates
a symmetric session key to encrypt the request that contains
Alice's login credentials.

The encrypted request is forwarded to \url{bank.com}'s CDN and
then to its origin server. Because the request is encrypted end-to-end
between the client proxy and the server proxy, any on-path adversary
inside the CDN cannot peek inside and acquire Alice's login
credentials. Similarly, the server proxy encrypts the private
content in \url{bank.com}'s responses to Alice.

Next, we describe each of \name's components and how they interact
with each other.

\subsection{Key Distribution \& Management}

A website's new public key is the root of trust in the \name
design. \name uses the existing DNS infrastructure to distribute the
new public key. A website stores its new public key in a TLSA record,
following the specification of DANE~\cite{RFC6698-TLSA}. Since a
website will not change its public key frequently, it can set a long
Time-to-Live (TTL) value for the TLSA record, \eg, on the order of
hours, to reduce the query load on DNS.

A website should use DNSSEC~\cite{RFC4033-DNSSEC} to protect the
integrity of the TLSA record. Without DNSSEC, an on-path active
adversary may tamper with the public key distributed in the TLSA record,
compromising \name's security.  According to a measurement study in
2019~\cite{IMC19-DNSSECRegistrar}, 15 out of 20 most-used domain
registrars support DNSSEC and all top-level domains such as
\url{.com} and \url{.org} are DNSSEC-enabled~\cite{TLDofDNSSEC}.
Hence, a website that desires to protect its users' privacy can find a
suitable domain name registrar or DNS provider that supports DNSSEC.

Despite an increasing number of DNSSEC-enabled domain registrars~\cite{IMC19-DNSSECRegistrar}, 
a client's default DNS resolver
may not be DNSSEC-enabled. Therefore, in \name, the client proxy and
the integrity verifier will send a query for the public key in
DNS-over-HTTPS (DoH)~\cite{RFC8484-DoH} to a DNSSEC-enabled resolver,
as shown in Figure~\ref{fig:architecture}. Currently, all mainstream
browsers can launch DoH requests, as they are essentially HTTPS
requests. A website can specify a trusted DNS resolver that supports
DNSSEC in its client proxy. \Name refers to TLS Encrypted Client Hello (ECH) for distributing public keys by DNS~\cite{TLSECH}. \Name optimizes ECH by adopting DoH to ensure that a DNSSEC-enabled resolver is used even when the client's default resolver does not support DNSSEC.

We note that such a key distribution mechanism will not heavily
increase the load of the DNS infrastructure nor introduce DDoS vulnerabilities because the request for the public key is a one-time cost. \Name caches
a website's public key in the browser and does not request for the public key until the user clears the cache.

\Name stores a public key instead of a certificate in the TLSA
record for two reasons. First, certificate verification requires a
client proxy to access the trusted root certificates of the client's
OS or browser. Currently, JavaScript code in a browser cannot
access those certificates. Supporting this design requires modifications to browser implementations. Second, a website can publish or revoke the public key independently and efficiently. 
It does not communicate with Certificate Authorities (CAs) for new certificate application or potential revocation, which may be costly and time-consuming~\cite{liang2014https}.
Our design simplifies a website's key management.



There are two alternative ways to key distribution. One is to
piggyback a website's public key on an HTML file delivered to the
client. However, this approach only works in the presence of a passive
adversary, since an active adversary in a CDN may replace the public key
in the HTML file. Another approach is for the client to bypass a
CDN and fetch the public key from the origin server directly, but it
exposes the origin server to DDoS attacks.



\subsection{The Client Proxy}\label{Design:ClientProxy}

In \name's design, the client proxy is a JavaScript library that
runs inside a web browser. A web service that desires to deploy \name
distributes this proxy to its users via a landing page (\eg,
\url{login.html} or homepage).  We design the client proxy to take a
configuration file as an input so that each website only needs to
configure the file to deploy \name and does not need to develop new
JavaScript code.  The client proxy is in charge of 1) querying DNS to download the web service's
new public key, 2) establishing an encryption channel with the origin
server, 3) encrypting a user's private requests sent to the server, and 4) decrypting the server's responses
that include private content.

A website customizes a configuration file (\texttt{configure.js}) in the client proxy
to include the URLs of private content and the handshake API. A client proxy uses the
handshake URL to establish an encryption channel with a server proxy.
A website can specify the set of private request URLs using regular
expressions to reduce the configuration overhead and the size of the
configuration file. An example of \texttt{configure.js} is shown in
\ifx\EXTEND y
    Appendix~\ref{Appendix:Implementation}.
\else
    the extended version of this paper~\cite{InviCloak-arxiv}.
\fi
We evaluate the configuration effort in \S~\ref{ConfigurationEffort}.



The main difference between \name's client proxy and the usual
JavaScript code is that \mr{the client proxy is a Service Worker~\cite{ServiceWorkerAPI},
    which runs in the background of a browser and works
    across different web pages from the same domain.}
A CDN-visible landing page installs the proxy by
registering it with the browser in the HTML, and the browser caches it
subsequently. When the browser navigates to a private web page in the
same domain, the client proxy does not need to be installed again.

\subsection{The Server Proxy}\label{Design:ServerProxy}
We design the server proxy to be an add-on module to a web server such
as NGINX~\cite{reese2008nginx}. A website need not modify its web
server's implementation to deploy \name. The website lists the
URLs of private content in the server proxy's configuration file (\texttt{nginx.conf}), 
similar to how it configures the client proxy.
An example of \texttt{nginx.conf} is shown in
\ifx\EXTEND y
    Appendix~\ref{Appendix:Implementation}.
\else
    ~\cite{InviCloak-arxiv}.
\fi
The server proxy is in charge of 1) establishing an encryption
channel with a client proxy, 2) decrypting requests encrypted by the
client proxy, and 3) encrypting private content returned to the client.

\subsection{Establishing the Encryption Channel}\label{Design:Establish}
We now describe how a client proxy establishes an encryption channel
with a server proxy.  The client proxy will first send a DoH query to
obtain the TLSA record that stores a website's new public
key. Then it will establish a secret session key with the server proxy for encryption and decryption. Our key exchange protocol is based on
TLS~1.3~\cite{RFC8446-TLS1.3}, as TLS security has been carefully
examined by researchers.
We describe the protocol in
\ifx\EXTEND y
    Appendix~\ref{Appendix:Handshake}.
\else
    an extended paper~\cite{InviCloak-arxiv}.
\fi

\textbf{Session key reuse:} We design the session key to be reused
across multiple requests/responses so that the session key setup
is a one-time cost in a web session.
According to TLS~1.3~\cite{RFC8446-TLS1.3}, \name will randomize a 32-byte session ID after each key exchange and store the corresponding session key. The session ID is returned to the client.
When the client proxy sends encrypted private content to the server proxy, it will include the session ID in plaintext.
The server proxy retrieves the session key through the session ID, and it will reject the request if no corresponding session key is found.


\textbf{On-demand vs. Asynchronous key exchange:} The key
exchange process can happen either on-demand or asynchronously.
Each approach has its pros and cons.
For on-demand key exchange, a client proxy triggers the key exchange when
it sends the first private request in a web session. For asynchronous key exchange, the client proxy starts the key exchange right after the user loads the landing page that distributes the client proxy. For example, the key exchange may happen when a user visits a website's login page. When the user
enters her login information, the session key is already available for
encrypting her password. In contrast, with the on-demand approach, the
key exchange happens when the user hits the password submit
button.

The advantage of the on-demand approach is that it does not
waste a website's resources to perform the key exchange in case the
user does not log in; the disadvantage is that it adds a round trip
time (RTT) to the load latency for the first encrypted request.
In contrast, the asynchronous approach shortens the page load time (PLT)
when a user requests private content for the first time.

We leave it as a configuration option for a website to choose between
on-demand or asynchronous key exchange.
We note that even with the on-demand approach, the key setup
is a one-time cost for an entire \name session.

%

\subsection{Using the Encryption Channel}\label{Design:EncrytionChannel}

After the session key setup, the client and server proxy can use it to
encrypt and decrypt private communications between a browser and a
website's origin server. These operations are described below.


\textbf{Encrypting an HTTP request:} When a browser initiates
a request whose URL is in the list of the client proxy's configuration
file, the client proxy generates a message that includes a session ID, a sequence number, and the request content.
The session ID uniquely identifies the \name session it shares with the server
proxy, and the sequence number uniquely identifies the request within this
session to prevent replay attacks (described shortly). The client
proxy encrypts the sequence number and the request content with the session key it shares with the server proxy, using a ciphersuite approved by TLS~1.3. It then sends the encrypted request to the CDN, which will forward it to the origin server.
\Name does not change TLS, so the underlying TLS connections
are unaware of the additional encryption added by the client proxy and
will treat the encrypted request as a regular HTTP request.

Our current design only encrypts the body of an HTTP request, because
most headers do not contain private content. The \texttt{Cookie}
header may contain important authentication information, and we
describe cookie encryption and management in
\S~\ref{Design:CookieManagement}. The current design can be extended
to encrypt other request headers specified by the client proxy's
configuration file.

\textbf{Client-side session state:} The session state a client
proxy maintains include the session ID, the latest outstanding sequence number, and the session key it shares with the server proxy.
The client proxy stores the session state
in a browser cache. Thus, an \name session may be valid across multiple
web sessions. For private requests to the same server, the
client proxy can reuse the session key. Web developers can configure an
expiration period for the sessions.

\textbf{Processing an encrypted HTTP request:}
When a server proxy receives an encrypted request, it uses the session ID
and the encrypted sequence number to detect and
prevent replay attacks. The server proxy maintains a sliding
window of recently received sequence numbers, and any out-of-window requests or
requests with duplicate sequence numbers are discarded by the server proxy.
For a valid request, the server proxy decrypts it, strips off
the additional fields added by the client proxy, and forwards the
decrypted request to the website's origin server.

The difference between the sliding windows of \name and TCP is that \name accepts an out-of-order request before its previous ones are accepted since a browser will launch HTTP requests in parallel.

\textbf{Encrypting an HTTP response:} When a server proxy receives an HTTP response of a
private URL from the origin server, it will attach the corresponding sequence number
to the beginning of the response body and then encrypt the response body with the session
key. The server proxy also encrypts
certain cookies specified in the \texttt{Set-Cookie} header to avoid
cookie leakage as described in
\S~\ref{Design:CookieManagement}.

\textbf{Decrypting an HTTP response:} When the client
proxy receives an encrypted response from the server proxy, it will
decrypt it using the session key. The response will include the
original sequence number to prevent a response replay attack. If the client
proxy finds that the sequence number matches the one it sends out, it
returns the response to the web browser; otherwise, it will discard the response.


%
%


%

\subsection{Integrity Verifier}\label{Design:IntegrityVerifier}
The client proxy and the server proxy are sufficient to defeat a
passive adversary residing inside a CDN, as the private communications
will be protected by \name's encryption channel.
However, an active adversary residing in a CDN may inject malicious code and
hijack the encryption channel, thereby gaining access to the private
content.

\Name uses integrity verification to defend against such an active
adversary. A website will use the private key corresponding to the
public key distributed in its DNS TLSA record to sign CDN-visible
objects such as the client proxy, HTML files, and JavaScript code.
\mr{We refer to these objects as \emph{CDN-visible executable objects},
    which also include the code of the client proxy. }
The other CDN-visible objects such as CSS and images are not executable,
and modifying them cannot break \name's protection. A website need
not sign them to reduce the computational cost. A website can sign the
CDN-visible executable objects offline before a CDN caches them using
any signing tool, such as OpenSSL~\cite{OpenSSL}.

We introduce an integrity verifier at the client side to verify the
signatures. Unlike a client proxy, the integrity verifier is a
browser extension to an existing browser. It is a component
independent of a web service and can be securely obtained outside a web session (\eg, via the extension market in a browser) as we assume in \S~\ref{Assumptions}.


When a request is sent out, the integrity verifier simultaneously
fetches the server's new public key through DoH (if not cached),
and it determines the \name's enablement of a website through
the existence of the TLSA record.
When a browser receives the response from the CDN, the integrity
verifier will intercept the response and examine whether the response
body is a CDN-visible executable object. If so, it validates the
signature of the object using the server's new public key.
If one of the web objects fails on the verification, the extension
will block the loading page and send a pop-up to
alert the user. The verifier can also prompt the user to
report the incident to a central repository. This repository could be
maintained by the distributor of the integrity verifier or any
third party that is interested in aggregating user reports to
collectively detect ongoing security threats. The integrity verifier
does not verify the encrypted responses from the server
proxy. Instead, the client proxy decrypts and validates the integrity
of such messages.

An alternative way to preserve the integrity of web resources is to use the existing web techniques, Subresource Integrity (SRI)~\cite{SRI} or Signed HTTP Exchanges (SXG)~\cite{SXG}. However, both SRI and SXG attach the integrity of subresources to the trust of the base HTML files, and they do not ensure such HTML files' integrity. Besides, they require extensive modification of HTML files, while \name is designed to ease the deployment by minimizing modification to existing web resources.

\subsection{Partial Deployment of Integrity Verifier}\label{Design:PartialDeployment}

%

In the \name design, a website can deploy the server proxy and the client
proxy without a user's involvement to defeat a passive adversary, but
the user needs to install the integrity verifier herself to defeat
an active adversary.

However, a partial deployment of the integrity verifier can
deter active attacks. This is because an adversary cannot tell which
user has installed the integrity verifier. If it launches an active
attack such as the code injection attack, it risks being caught by
those users who have installed the integrity verifier. Admittedly, an
adversary may attempt to conduct a targeted attack, \eg, modifying the
client proxy code for those users who have not installed the integrity verifier.
\mr{
    A typical method to learn about a client's information is to use the \texttt{User-Agent} header, but \name can be configured to mask this header in the client proxy to prevent such information leakage.
}\mr{We acknowledge that some
    browsers do not support browser extensions currently, but
    overall, \name provides an option for security-conscious users and websites to protect their sensitive data from potential active attackers by using the browsers that support browser extensions.}

\subsection{Cookie Management}\label{Design:CookieManagement}
A website may issue a cookie to a user after the user successfully
authenticates herself. We refer to such a cookie as a \emph{user
    authentication cookie}. When a user's HTTP request presents the cookie,
the website may return private content available only to an
authenticated user. Currently, such cookies are visible to a CDN in an
HTTP request/response header. An adversary inside a CDN may intercept
this cookie and attempt to use it to access a user's private content.

\Name prevents such leakage by encrypting a user's authentication
cookie with the session key established between the client and the server
proxy. A website will specify which cookies are private to a CDN
(referred to as \emph{private cookies}) in its server proxy's configuration file.
The server proxy will encrypt the
private cookies in the \texttt{Set-Cookie} header of an HTTP response
and decrypt the cookies in the \texttt{Cookie} header of an HTTP
request. A client proxy need not decrypt the cookies, so the browser
will store the encrypted ones and attach them to request headers when
needed.

The encryption binds a user authentication cookie to the \name
session where the session key is established. When an
on-path adversary replays an encrypted user authentication cookie, it
cannot impersonate the user, because it does not have the
corresponding session key to generate a valid private request to
the server. It cannot decrypt an encrypted response, either.
Therefore, the adversary cannot gain access to a user's private
content.

Since \name's session state are stored in the browser cache,
\name can decrypt the existing cipher-cookies as long as a user does not
clear the cache, and the \name session is not expired.
Thus, a user does not need to re-login when she revisits a web page.


We note that websites may share cookies with CDNs. For example, a
website may share a paywall cookie after a user logs into a paywall
and obtains content behind the paywall. Such cookies are separate from
authentication cookies and are referred to as \emph{signed
    cookies}~\cite{SignCookieAmazon, SignCookieGoogle}. Signed cookies
should be configured as CDN-visible and \name will not encrypt them
nor affect the existing cookie-sharing practice between a CDN and a
website.





\section{Security Analysis}\label{SecurityAnalysis}

In this section, we analyze \Name's security properties. Our analysis
suggests that \name can preserve confidentiality and integrity in the presence of a compromised CDN or a compromised DNS provider.


\textbf{Man-in-the-Middle attacks:}
An attacker inside a CDN is on the communication path between a client
and an origin server and may attempt to eavesdrop and tamper
the messages it transmits. However, \Name enables a client and
an origin server to establish a secret session key to encrypt their
private content. A server signs its key exchange message with a
private key that it does not share with the CDN, so the CDN cannot
compromise the session key setup process. Therefore, even if an on-path
adversary may have access to the TLS private key of a website, it
cannot eavesdrop or tamper with the private content that a client and
an origin server exchange.

\textbf{Code injection attacks:} In the \name design, the client
proxy is a JavaScript library, so a client may download it from a
CDN. An attacker inside a CDN may attempt to inject or modify the code
in the client proxy to obstruct \name's session key setup.
\Name uses the integrity verifier (\S~\ref{Design:IntegrityVerifier})
to prevent such attacks.


\textbf{Replay attacks:} An attacker inside a CDN may replay
messages it receives between a client and an origin server. We prevent
these attacks by including a session ID and an encrypted sequence number
in a client's request (\S~\ref{Design:EncrytionChannel}).
The server proxy can use the request ID to detect a replayed client request.

%
%
%

\textbf{Forward secrecy:} \Name provides forward secrecy because it
uses the Diffie-Hellman key exchange protocol as in TLS
1.3~\cite{RFC8446-TLS1.3} to encrypt private content.
The long-term secret in the protocol is the website's private key.
Even if this private key or future session keys are
compromised, an attacker cannot decrypt the messages sent in the past sessions.



\textbf{Impersonating as a user:} As in TLS~\cite{RFC8446-TLS1.3},
\name's key exchange protocol does not authenticate a user. A
malicious adversary may attempt to impersonate as a user to access
private content, but we hypothesize that a website will use additional
authentication mechanisms such as a login password to protect private
content. Therefore, although the adversary can establish a session key
with the origin server as any client can, it cannot authenticate as
the user. In addition, we encrypt the user authentication cookie
using the session key after a user authenticates herself. As described
in \S~\ref{Design:CookieManagement}, this design binds a user's
authentication cookie to the \name session key. Thus, an on-path attacker cannot impersonate a user
to access the user's private content protected by the session key.



\textbf{DDoS attacks:} \Name does not change the current CDN
service model, so an origin server enjoys the same DDoS protection
benefits as in the current service model. The current DDoS protection
service of a CDN has the caveat that if an attacker uncovers the IP
address of the origin server~\cite{vissers2015maneuvering}, she may
directly flood DDoS traffic at the server. \Name provides an additional
benefit in this case. Since a client will only communicate with an
origin server via a CDN, the server can use router filters that 
are resistant to source address spoofing to whitelist the traffic
from the CDN~\cite{Netfilter,ehrenkranz2009state}.




\textbf{Application-layer attacks:} A CDN may act as a Web
Application Firewall (WAF) to filter client requests that contain
application-layer attack payloads~\cite{gillman2015protecting},
including SQL injection~\cite{halfond2006classification}, cross-site
scripting (XSS)~\cite{weinberger2011systematic}, and application-layer
DDoS attack requests~\cite{xie2008monitoring,ATC19-Finelame}. With
our solution, a client's requests that include private content are
encrypted. Although a CDN can continue to filter unencrypted requests,
it can no longer filter encrypted requests.

A website can employ a WAF by itself to defend against such attacks.
There exists an open-source WAF, ModSecurity~\cite{ModSecurity},
available to two commonly used web servers: Apache~\cite{Apache} and
NGINX~\cite{reese2008nginx}. Besides, the WAF rules for commonly known
attacks are openly available~\cite{CRS}. Developers can maintain a WAF
with NGINX and ModSecurity through a simple one-line configuration.
Moreover, a general defense provided by a CDN might be ineffective,
as application-layer attacks such as algorithmic complexity DDoS
attacks~\cite{crosby2003denial,ATC19-Finelame} can be site-specific. A
website needs to deploy a site-specific WAF anyway for effective
attack protection. Given \name's protection for privacy, a website may
consider it an acceptable tradeoff to deploy an on-site WAF and filter
malicious requests after they are decrypted.

\textbf{A compromised DNS provider:} \name uses DNSSEC and DoH to
secure key distribution. A website's public key may be tampered with when
its DNS provider is compromised. In this case, as long as the
adversary does not obtain the website's TLS private key or session key
shared with the website's CDN provider, it cannot successfully
impersonate a website to complete the TLS exchange or the session key
exchange between a client proxy and a server proxy. Therefore, the
adversary is still unable to gain access to users' private content.

A powerful attacker that compromises both a website's DNS provider and
CDN provider will compromise \name's security. Technically, it is
possible to defeat such threats by distributing a new TLS certificate
instead of a public key in a website's TLSA record and modifying
browser implementations to validate the new certificate. However, we
believe the risk of such threats is low and opt for a design that
does not require browser modifications.

%
%

\section{Implementation}~\label{Implementation}
We implement a prototype client proxy using JavaScript and Service Worker
API (SW)~\cite{ServiceWorkerAPI}.  The same implementation works for
browsers that support SW including Firefox, Chrome, Safari and Edge.
\mr{SW has been enabled in mainstream browsers (Firefox, Chrome, Safari,
    and Edge) since 2018~\cite{ServiceWorkerReady}. In the case where a
    browser is outdated and does not support SW, the server proxy will
    reject the request without encryption by SW for security reasons and
    return a response to prompt the user to upgrade her browser.}

We use the Web Cryptography API~\cite{WebCryptographyAPI} provided by browsers for cryptographic algorithms. We use the ciphersuite AEAD-AES256-GCM-SHA384 for symmetric encryption, and use the curve NIST P-256 for Diffie-Hellman key exchange. Both algorithms are approved by TLS1.3~\cite{RFC8446-TLS1.3}.
The lines of JavaScript code for the client proxy code (excluding
the configuration file) is $\sim$590.

For the integrity verifier, we implement it as a Chrome and Firefox
extension. It takes less than 300 lines of JavaScript code.  \mr{ One
    complication is that the current implementation of Chrome lacks a
    browser extension API, \texttt{webRequest.filterResponseData()}, to
    read the response body~\cite{ChromiumBugs}, which is already
    implemented in Firefox~\cite{FirefoxWebRequest}. A discussion and a
    proposal from Chrome developers show that the API does not introduce
    new vulnerabilities, and it is missing because of the technical
    complexity\cite{WebRequestProposal,WebRequestGroups}.  } Thus, we
provide a Chromium (the open-source version of Chrome) patch to
implement the function.  

For the server proxy, we implement it in C as a module to a popular web server NGINX.
It takes $\sim$2000 lines of code. We release our code to facilitate \Name's deployment~\cite{InviCloakCode}.
We include a more detailed description of our implementation and sample configuration files in
\ifx\EXTEND y
    Appendix~\ref{Appendix:Implementation}.
\else
    an extended paper~\cite{InviCloak-arxiv}.
\fi


\section{Evaluation}\label{Evaluation}
We evaluate the performance of our prototype implementation of \name,
and compare it with related work. First, we micro-benchmark the
computational overhead by measuring the computation time of each \Name operation.
Second, we evaluate how the
added overhead affects an origin server and a CDN's edge server's
throughput along with the page load times (PLTs) of
realistic web pages at the client side. We compare this overhead with the TEE-based solution, Phoenix~\cite{Security20-Phoneix}.
We also evaluate \name's overhead on a modern web application using Cloudflare.
Besides performance overhead, we estimate how much
effort it takes to deploy \name at a website using the lines of
configuration a website needs to make to deploy \name.
Finally, we analyze the deployment cost of \name and compare it with
Phoenix. We do not directly compare \name's performance with the
TLS-modification-based solutions~\cite{NDSS19-maTLS,SIGCOMM15-mcTLS,SP18-formal} and the two-domain solution~\cite{NDSS16-CDNonDemand} as their performance can be approximated by the
baseline client/server performance without \name enabled~\cite{SIGCOMM15-mcTLS,NDSS16-CDNonDemand}.
We describe each of the experiments and the evaluation results in
detail.

\textbf{Testbed:} We set up a small testbed of three
Dell Precision T3620 machines in our experiments.
The three machines serve as a client, a CDN, and a website's
origin server, respectively.  Each machine has an Intel Core i7-7700
CPU and 32~GB of RAM, and runs Ubuntu 16.04. The three machines connect
to each other via Ethernet. We use the tool, Linux Traffic Control,
(\texttt{tc})~\cite{hubert2002linux} to configure the bandwidth and
RTT values between the machines if necessary. The server machine then runs the NGINX
implementation and configuration as we describe in
\S~\ref{Design:ServerProxy}. In order to emulate CDN functions, we set up an NGINX proxy or Phoenix
at the CDN machine. In our experiments, we use Chromium (Version 87.0.4280.88) as the browser.


\subsection{Computation Overhead}
\textbf{Experiments:} We let the client machine load our client proxy implementation into Chromium. It sends synthetic web requests to the
origin server, and the server responds with synthetic web responses.
In order to measure how the payload size affects the computational overhead, we vary the payload size of an HTTP request/response from 2~KB to 8~MB, because 90\% of web pages are smaller than (8~MB)~\cite{HTTPArchive} at the time of this work.

We instrument the client proxy, the server proxy, and the integrity verifier to record the computation time of the
encryption, decryption, and signature verification operations.
We do not show signature generation overhead because websites can sign the static files offline.
The measurement includes the complete computational overhead caused by the cryptographic algorithms and the other necessary operations such as memory allocation and copy.
We do not measure the session key setup overhead in this experiment, as it is a
one-time cost per session. However, we do measure it in the PLT experiment
(\S~\ref{ClientPageLoadTime}).  We repeat each experiment one hundred
times and use the mean values as results.

\begin{figure}[tb]
    \centering
    \subfloat[\textbf{\small Client side}]{
        \begin{minipage}[t]{0.49\columnwidth}
            \centering
            \includegraphics[width=\textwidth]{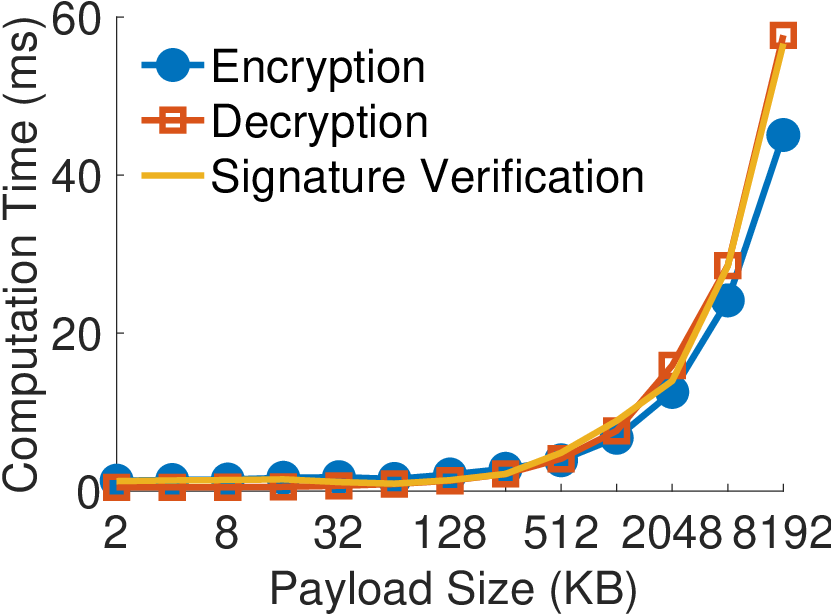}
        \end{minipage}\label{fig:clientside}
    }
    \subfloat[\textbf{\small Server side}]{
        \begin{minipage}[t]{0.49\columnwidth}
            \centering
            \includegraphics[width=\textwidth]{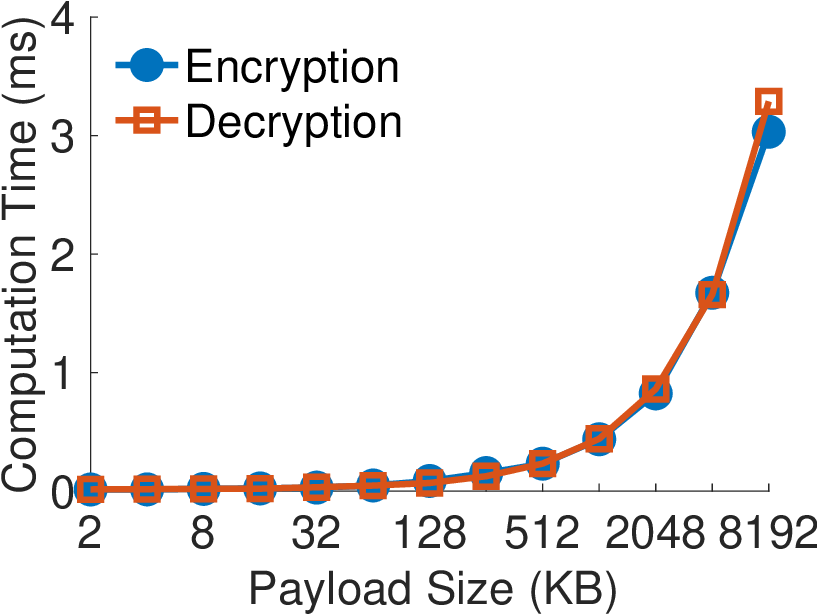}
        \end{minipage}\label{fig:serverside}
    }
    \caption{\label{fig:CT}\textbf{\small This figure shows the computational time our testbed machines take to perform \name's cryptographic operations.}}
\end{figure}

\textbf{Results:} Figure~\ref{fig:CT} shows the results of these experiments.
The encryption, decryption, and signature overhead grow
almost linearly with the size of the payload. The computation overhead
at the server is one order of magnitude smaller than that at
the client in all cases. The reason for this performance gap is that we implement
the server proxy in C, while the client proxy is in JavaScript for
easy deployment.
\mr{
    For payloads smaller than 8~MB, the encryption
    and decryption take around 3~ms at the server.
    At the client, the encryption takes less than 50~ms while the decryption
    takes around 60~ms. Decryption is slower than encryption at the client side because the  decryption code is more complicated
    in our implementation and  supports streamlined decryption of large responses.
    The client takes around 60~ms to verify the signature of an 8~MB file. Our test using a simple tool we developed for signature generating takes 16~ms for an 8~MB file on the server.
}

\mr{
    We consider the computation overhead of \Name acceptable at both
    the server side and the client side. Furthermore, as the median
    web page size is around 2~MB~\cite{HTTPArchive}, much smaller than 8MB, we expect
    \name's overhead to be low in practice.
    We validate this argument by evaluating the PLTs in \S~\ref{ClientPageLoadTime} and \S~\ref{InternetPageLoadTime}.
}

\begin{figure*}[t]
    \centering
    \subfloat[\small\textbf{CDN throughput}]{
        \begin{minipage}[t]{\columnwidth}
            \centering
            \includegraphics[width=0.92\textwidth]{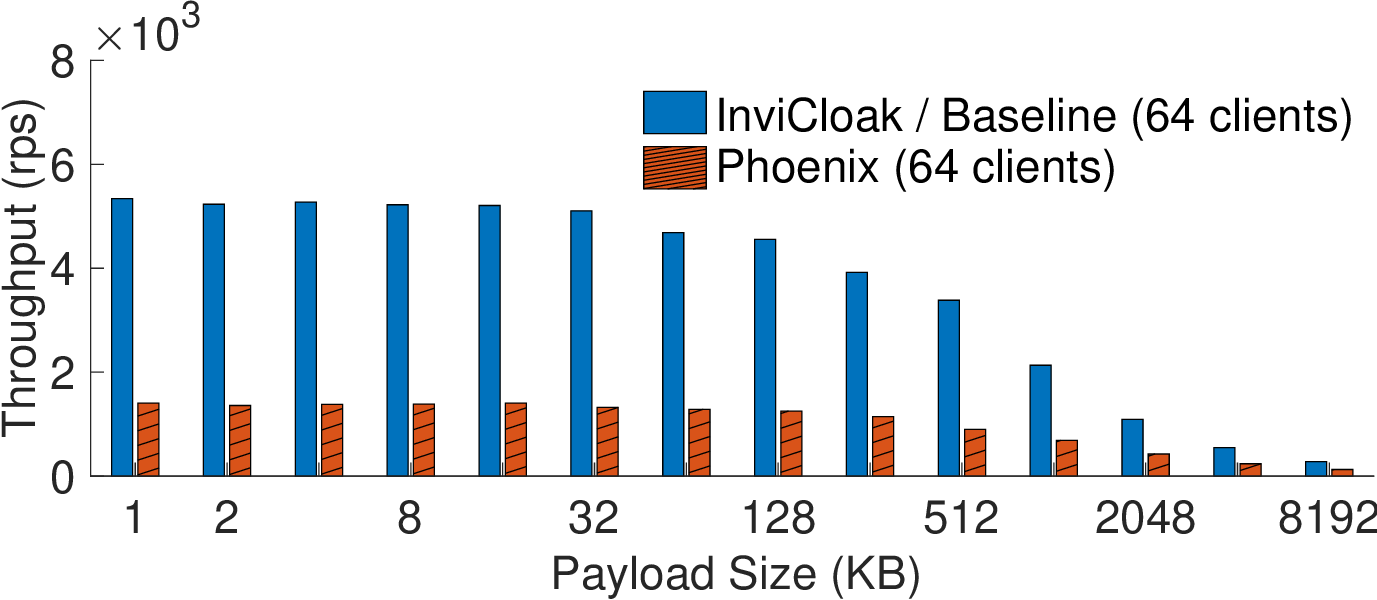}\label{fig:throughput-cdn}
        \end{minipage}
    }
    \hspace{0.05\columnwidth}
    \subfloat[\small\textbf{Server throughput}]{
        \begin{minipage}[t]{\columnwidth}
            \includegraphics[width=0.92\textwidth]{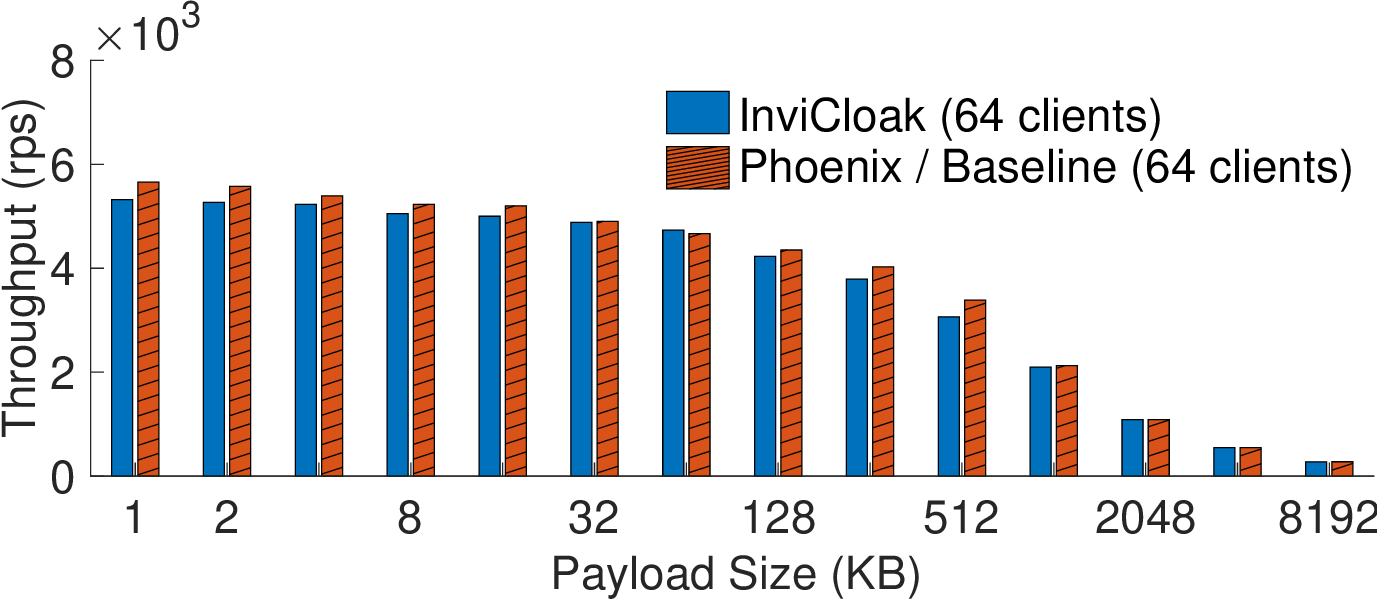}\label{fig:throughput-server}
        \end{minipage}
    }
    \caption{\small\textbf{This figure shows how \name and Phoenix
            affect a server and a CDN's throughput. The throughput is in
            the unit of HTTP responses per second.}}
\end{figure*}

\begin{figure}[t]
    \centering
    \subfloat[\small\textbf{Alexa websites}]{
        \begin{minipage}[t]{0.49\columnwidth}
            \centering
            \includegraphics[width=\textwidth]{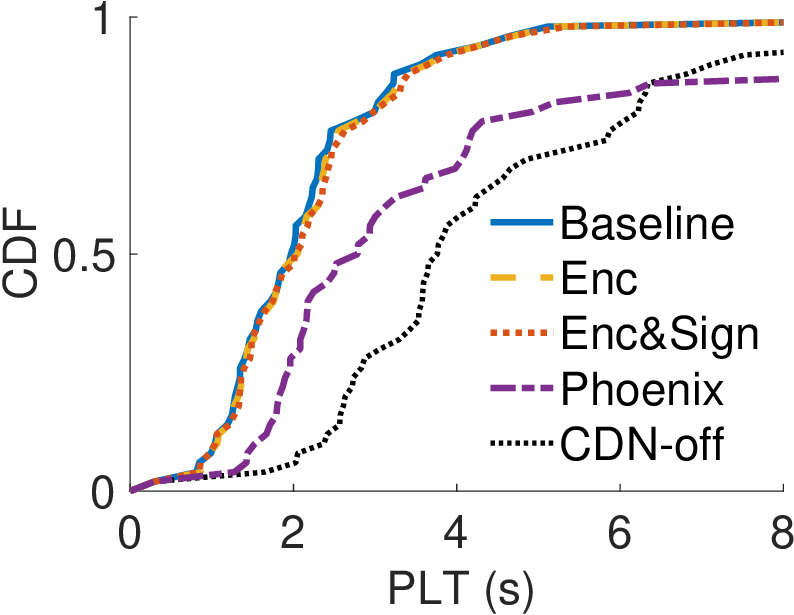}\label{fig:mockwebsites}
        \end{minipage}
    }
    \subfloat[\small\textbf{DeathStarBench}]{
        \begin{minipage}[t]{0.49\columnwidth}
            \includegraphics[width=\textwidth]{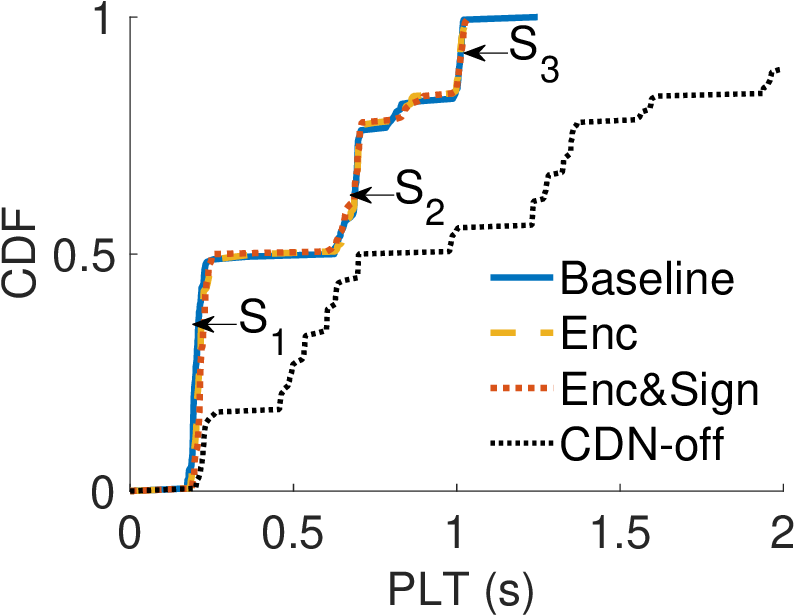}\label{fig:webapp}
        \end{minipage}
    }
    \caption{\small\textbf{(a) CDF of PLTs on 50 Alexa websites (\S~\ref{ClientPageLoadTime}). (b) CDF of DeathStarBench's PLTs from six geographically distributed clients (\S~\ref{InternetPageLoadTime}). The unit of each figure's x-axis is second.}}
\end{figure}

\subsection{CDN \& Server Throughput}\label{ServerPerformance}
\textbf{Experiments:} We run two sets of experiments to test
the throughput of the CDN and the server, respectively. We use
GoHTTPBench~\cite{GoHTTPBench}, which is a multi-thread version of
ApacheBench~\cite{ApacheBench}, to send 50K HTTP requests in total
from the client machine to the CDN machine or to the server
machine. \mr{We vary the size of an HTTP response from the CDN or the
    server from 1~KB to 8~MB. GoHttpBench will send requests with 64
    concurrent threads to simulate 64 concurrent clients,
    which keeps the server machine or the CDN machine's CPU usage exceeding 95\%.}

For the CDN performance evaluation, we measure the throughput of
Phoenix and a baseline (an NGINX proxy). Since \name does not require
any modification of a CDN, we use the baseline performance to
represent the CDN performance when \name is deployed.
For the origin server performance, we measure the throughput of an
NGINX proxy with \name and a baseline (an NGINX proxy without
\name). Since Phoenix does not require any modification of an origin
server, we use the baseline performance to represent the origin
server's performance when Phoenix is deployed.

\textbf{Results:} Figure~\ref{fig:throughput-cdn} shows the
results of the CDN performance experiments. As we can see, Phoenix
introduces considerable overhead to a CDN server. Its throughput is
only about one-third of baseline's in each experiment, which is consistent
with the results in the original Phoenix paper~\cite{Security20-Phoneix}.
Figure~\ref{fig:throughput-server} shows the results of origin server
performance experiments. \mr{\Name reduces the server throughput by less
    than 5\% for payload sizes ranging from 1~KB to 8~MB.} Overall, due to
different designs, \Name and Phoenix possess different
advantages. Phoenix preserves the origin server's throughput but
degrades the CDN's throughput significantly. \Name does not introduce
overhead to a CDN server, but slightly slows down the origin server.

\begin{table}[t]
    \caption{\label{tab:statistics}\textbf{\small The sum of object sizes in the crawled web pages.
            Login pages do not contain any private content. The numbers in parentheses indicate the number of URLs. `JS' refers to JavaScript.}}
    \begin{center}
        \begin{tabular}{l|l|l|l}
            \multicolumn{2}{c|}{}        & Median  & Mean                        \\
            \hline
            \multirow{3}*{Login pages}   & All     & 2476~KB (47) & 3895~KB (63) \\
                                         & HTML    & 85~KB (4)    & 225~KB (4)   \\
                                         & JS      & 1794~KB (17) & 2815~KB (25) \\
            \hline
            \multirow{4}*{Private pages} & All     & 5361~KB (74) & 6896~KB (93) \\
                                         & Private & 210~KB (7)   & 505~KB (17)  \\
                                         & HTML    & 212~KB (4)   & 344~KB (6)   \\
                                         & JS      & 3551~KB (30) & 5378~KB (34) \\
        \end{tabular}
    \end{center}
\end{table}

\subsection{Client Page Load Times}\label{ClientPageLoadTime}
We use realistic web pages to estimate how \name affects a user's
perceivable performance. We crawl the two groups of web pages from the top-50 websites that we are able to register and log in from Alexa's top-100 websites.
For the first group, we crawl the login page of each website. Some websites embed login forms inside the homepage instead of a separate login page. In such cases, we crawl their homepages as the login pages. For the second group, we manually log into each website and crawl one private web page that contains private information such as the profile, the visiting history, or the list of favorite items. We mirrored these web pages in our testbed.

For the login pages, we regard each web object as CDN-visible as it does not require a user login.
For the private pages, we use the HTTP \texttt{Cache-Control} header to
identify the non-cacheable web objects. We regard all non-cacheable objects as private objects in our evaluation and include their URLs in the configuration files.
\Name will encrypt such objects in the experiments.
For each web page, we sum up the size of embedded web objects.
Among the crawled web pages, we calculate the median and mean of the sum values,
as shown in Table~\ref{tab:statistics}.
The average size of login pages and private pages is 3895~KB and 6896~KB, respectively.
On average, a private page contains 505~KB of private objects.

To measure the PLT of each test web page, we instrument the browser at the
client machine send all requests to the CDN machine. The CDN machine
replies with the CDN-visible web objects it caches and forwards the
private requests to the server machine. To present a more realistic
evaluation, we limit the bandwidth between machines to
100~Mbps. According to \cite{CDNRTT}, the average RTT between clients
and Cloudflare's edge servers is 36~ms. Thus, we use this value as the
RTT between the client and the CDN machine. We set the RTT between the
CDN and the server as 100~ms, which is approximately the RTT measured
between two machines located on the east coast and the west coast of
the United States. We load each web page five times in Chromium and
compute the average PLT for each web page.

We evaluate \name's initiation overhead by measuring the PLT increments
of the crawled login pages when they are used as \name's landing page.
The initiation process includes the verification of CDN-visible executable objects, installation of the client proxy, and the asynchronous key exchange as discussed in \S~\ref{Design:Establish}.
The results show that \name's initiation increases a login page's PLT by less than 8\% for all websites.
Specifically, the overheads on 47 websites are less than 5\%. We note that such an overhead is a one-time cost, as the CDN-visible objects and \name's session state will be cached by the browser after a user visits the landing page for the first time.

As for the crawled private pages, we run experiments in multiple
settings to show how various computational overhead affects the PLTs.
For the first setting, we measure the PLTs when both \name and Phoenix are
disabled as the baseline (\texttt{Baseline}).
In the second setting, the client machine runs only the client proxy
without the integrity verifier (\texttt{Enc}). For the third one,
the client machine runs both the client proxy and the integrity
verifier (\texttt{Enc\&Sign}).  In these two settings of \name,
the client proxy has already set up the session key with the server proxy
in the landing page.  To compare \name with the existing solution, we
measure the PLTs when Phoenix is enabled (\texttt{Phoenix}).
\mr{All the settings above use a CDN with warm cache.
    We also measure the PLTs with cold cache, \ie,
    the CDN cache is disabled (\texttt{CDN-off}).}


Figure~\ref{fig:mockwebsites} shows the results of the 50 websites in
each setting.  \mr{The PLTs under the \texttt{Baseline}, \texttt{Enc},
    and \texttt{Enc\&Sign} settings have similar distributions, and
    there is a large gap between them and the \texttt{CDN-off} curve.
    This result suggests that \name largely preserves a CDN's
    performance benefit.}
Besides, \name's overhead is lower than that
of Phoenix. More than 32\% of PLTs exceed 4.0s when Phoenix is on, but
more than 92\% of PLTs under the \texttt{Baseline}, \texttt{Enc}, and
\texttt{Enc\&Sign} settings are less than 4.0s. The median PLT with
Phoenix is 2.8s, while the median PLTs for \texttt{Baseline},
\texttt{Enc}, and \texttt{Enc\&Sign} are 2.0s, 2.1s, and 2.1s, respectively.

%

\subsection{A Modern Web Application with Cloudflare }\label{InternetPageLoadTime}
\begin{mrlong}
    Evaluating \Name in a laboratory setting with crawled web pages may not
    fully capture \Name's performance in reality.
    To address this limitation, we run a modern web application, DeathStarBench~\cite{ASPLOS19-DeathStarBench},
    and evaluate \Name with geographically distributed clients on the Internet. We use Cloudflare, one of the largest CDN providers
    as the CDN.
    We cannot compare with Phoenix in this evaluation because TEE is not available on any CDN today.

    DeathStarBench is developed as a benchmark for studying the performance of modern cloud services~\cite{ASPLOS19-DeathStarBench}.
    It adopts the popular microservice architecture and is used by several
    academic and industrial institutions~\cite{ASPLOS19-DeathStarBench}. We use the
    social network provided by DeathStarBench in our evaluation, which provides functions
    similar to Twitter.
    In addition to a realistic web application, we also use realistic web content for our evaluation.
    We import a small social graph with 962 nodes and 18,812 edges from Facebook into the
    social network~\cite{AAAI15-socfb}. We also crawled 612,455 tweets from
    randomly selected 962 Twitter users through Twitter API~\cite{TwitterAPI}, and imported the tweets into the database.

    We deploy DeathStarBench with the crawled workload on a virtual machine of AWS in Virginia. We use six
    geographically distributed virtual machines of AWS as the clients. The clients are
    distributed in six AWS regions, including California, Montreal, Paris, Singapore, Sao Paulo, and Bahrain. Each virtual machine used in this experiment has 4 vCPUs and 8~GB of RAM and runs Ubuntu 20.04.  We use Cloudflare as the CDN to cache the static resources near the clients.

    In our experiment, we randomly select 10 users from the database, and instrument each
    client to login as each user and visit three private pages of the user. The three
    pages include a page showing at most 100 recent tweets of a user's followees,
    a page showing at most 100 recent tweets  of the user,
    and a page showing the user's all followees and followers. To show a
    conservative result, we regard all tweets, the followee list, and the follower list
    as private data. They are encrypted and not cached on Cloudflare. The other
    resources are public and cached on Cloudflare. To be conservative, we not only sign HTML and JavaScript files, but also sign CSS files.
    As defined in \S~\ref{ClientPageLoadTime}, we run experiments in four settings: \texttt{Baseline},
    \texttt{Enc}, \texttt{Enc\&Sign}, and \texttt{CDN-off}.
    Overall, each of the six clients will visit each of the three pages of all 10 users
    for five times. We compute the average PLT for each page.
    Thus, we have $6 \times 3 \times 10=180$ PLT values in each setting.

    As shown in Figure~\ref{fig:webapp}, we can find three explicit stages in the curves: $S_1,S_2,S_3$, which originates from the variance of PLTs between different locations. The low network latency in California, Montreal, and Paris lead to small PLTs shown in $S_1$. With the larger network latency, $S_2$ includes the PLTs in Sao Paulo and Bahrain, while $S_3$ shows the PLTs in Singapore. Besides stages, a large gap exists between the curve of \texttt{CDN-off} and the other ones, indicating that
    \name retains the performance benefit of a CDN.
    Moreover, \texttt{Baseline}, \texttt{Enc}, and \texttt{Enc\&Sign} have similar distributions,
    indicating \name has low overhead. Specifically, in either setting of \name,
    \name introduces less than 5\% overhead to 170 of all 180 PLT values.
    Therefore, we conclude that \name's performance overhead is low.
\end{mrlong}

\subsection{Deployment Effort}\label{ConfigurationEffort}
In this section, we estimate \name's deployment effort.

We use the lines of code (LoC) to measure the configuration overhead of
crawled websites and DeathStarBench.
For the 50 websites, we add the private URLs of each site into the
server proxy configuration file (\texttt{nginx.conf}) and edit the client proxy configuration file (\texttt{configure.js}) to include the private URLs.
We insert the code for Service Worker registration into each website's login page.
The LoC of \texttt{nginx.conf} ranges from 4 to 62 with an average of 13.7 and a median of 8.0. For \texttt{configure.conf}, the LoC ranges from 10 to 68 with an average of 19.7 and a median of 14.0. For all websites, the LoC of registration is 4.
\mr{
    As for DeathStarBench, we also use 4 LoC to register a Service Worker in the login page. Besides, we add a \texttt{configure.js} with 9 LoC and inserting 14 LoC into the existing \texttt{nginx.conf}.
}

\begin{mrlong}

    \Name's deployment requires a website developer to explicitly separate private URLs from public URLs. We discuss this process and the impacts of possible configuration errors made by a website's developer
    in two cases.



    First,  if the website developers have categorized the URLs into private and public directories (e.g. \url{/private} and \url{/public}), they can list the private URLs by wildcard or regular expressions in the configuration file. The configuration effort in this case is low. If the website developers erroneously categorize the URLs, the errors will be propagated into \name's configuration.

    Second, if the URLs are not categorized, we propose a method to automatically generate \name's configuration through the \texttt{Cache-Control} header as we did in the experiments of \S~\ref{ClientPageLoadTime}. This method is effective because current website developers manage the URLs that should not be cached through the HTTP headers~\cite{AzureCacheControl,CloudflareCacheControl}. With such a method, the configuration effort is also low.

    There are two sources of errors in this case: 1) developers  erroneously configure the \texttt{Cache-Control} header; 2) a website may serve non-cacheable public objects or cacheable private objects. The first source is introduced by a website's erroneous usage of the CDN and is not introduced by \name.
    As for the second source, our method would regard a non-cacheable public URL as private and a cacheable private one as public. The former only affects \name's efficiency.
    The latter will lead to privacy leakage. However, for websites that would like to protect user privacy, it is a misconfiguration to allow a CDN to cache private objects~~\cite{CloudflareCacheExplained,WCD,Security20-WCD}. Therefore, this type of error also originates from websites' URLs misconfiguration.


    In summary, the configuration effort of \name is low, as the configuration can be inherited from the existing website URL categorizations. As a result, any existing errors in a website's URL categorization may be propagated into \name's configuration and some of the errors may divulge user privacy.


\end{mrlong}




\subsection{Estimating Hardware Deployment Cost}\label{DeploymentCost}
We compare the monetary hardware deployment cost of Phoenix and \name by estimating the amount of
new hardware a CDN provider or a website needs to purchase.
\mr{We note that this estimate does not include the changes of operational costs when a website deploys \name or when a CDN upgrades to use TEE, as those numbers are difficult to obtain.}

Phoenix requires a CDN to support SGX, but it does not require any
upgrade on a website's origin servers. To make a lower bound estimate,
we assume that all existing edge servers of a CDN support SGX. Thus, a
CDN does not need to replace the old servers with SGX-enabled
servers. We take Akamai, one of the largest CDN providers, as an example to show the cost of the additional edge servers.

As shown in Figure~\ref{fig:throughput-cdn}, the throughput of Phoenix is about one-third of a CDN's edge server. Therefore, when
Phoenix is enabled, a CDN should at least add two times the amount of
existing edges servers to achieve a comparable throughput. Since Akamai has more than $290K$ edge servers spread around the world~\cite{rula2020s}, it needs to add about $290K \times 2=580K$ edge servers to maintain its current throughput. As each Intel Xeon processor that supports SGX will cost at least \$200~\cite{IntelProcessors},
the lower bound of the cost to Akamai is $580K\times\$200=\$116,000K$, which exceeds \$100M.



\Name is an end-to-end solution, so it does not require any
modification on CDNs. However, unlike Phoenix, it introduces computational overhead to a website's origin server.
To estimate the cost to upgrade for a website to deploy \name, we take an e-commerce website, \emph{Etsy}~\cite{Etsy}, a US top-50 sites, as an example in the estimate.


According to~\cite{zhang2013maygh}, Akamai observes 395~GB of traffic for Esty during a 18-hour period, namely 4909~KB per second.
Existing research shows that 74.2\% of requested bytes are
cacheable~\cite{zhang2013maygh}. Therefore, we estimate that the
origin server Esty responds with $4909\times \frac{1-74.2\%}{74.2\%}=1706$~KB
of payload per second. Note that
\name acts as a reverse proxy at the server side, so the added
overhead of \name is independent of the application's logic.
\mr{According to Figure~\ref{fig:throughput-server}, even when the payload size of each response is as small as 1~KB, an \name server proxy can serve more than $5000 \times 1=5000$~KB payload per second.}
Therefore, if we assume that a website like Etsy deploys the server proxy on a
separate machine, it only needs to add one additional machine as those in our testbed to
support its traffic load. The cost of each machine in our testbed is
less than $\$600$ at the time of this work.
Moreover, as the server proxy slows down an origin server's throughput by merely 5\% in our experiments,
if a website's origin server is not running near its full capacity,
a website may collocate the server proxy with its origin server without purchasing any new machine.
\mr{We omit the hardware cost of on-premise WAFs, because according to a survey~\cite{FiremonReport}, on-premise firewalls are popular among web services. More than 98.1\% already host on-premise firewalls.}



\mr{
    We note that this cost comparison is not an apple-to-apple comparison because of the essentially different designs of the two solutions. Upgrading a CDN with Phoenix will secure all websites that use the CDN, while upgrading a website with \name only secures the website itself.
    We do not sum up the cost of all websites using the CDN, because the costs of website upgrades are distributed among their corresponding organizations. Each organization makes an independent decision, enabling a gradual process of deployment.
}


\section{Limitations and Discussion}\label{Discussion}
\begin{mrlong}

    \textbf{Missing extension API in Chrome and insufficient extension support in mobile browsers:} The integrity verifier of \Name
    requires an extension API missing in Chrome~\cite{ChromiumBugs,WebRequestProposal,FirefoxWebRequest}.
    This missing API currently limits the extension's deployment on Chrome, but \name still provides a viable option for a website and a user who desire to protect their private communications.
    This is because both the desktop and mobile versions of Firefox already implemented the API~\cite{FirefoxWebRequest}. A user can switch to Firefox if she is concerned with active attacks.
    Moreover, we envision this work can facilitate the enablement of the extension in mobile Chrome and our implementation of the Chromium patch can assist in the implementation of the missing API in Chrome.
    Furthermore, the key idea of securely distributing the new key pair through the combination of DoH and DNSSEC in today's Internet is not limited by the extension. The integrity verifier can be implemented
    as a process separated from the browser.




    \textbf{Limiting CDN's WAF functions:}
    \Name encrypts a user's sensitive requests so a CDN can not apply WAF rules to
    those requests.
    We consider this limitation acceptable because websites that are concerned with privacy already built firewalls for private data on their origin servers. A Firemon report from 573 professionals in 2019 shows that 40\% of respondents have only on-premise firewalls, and 58.1\% of respondents host both on-premise and cloud firewalls~\cite{FiremonReport}. Therefore, 98.1\% of the surveyed professionals already hosted on-premise firewalls.
    It is possible that a website that uses cloud firewalls may need to extend its current on-premise firewall capabilities after it deploys \name, but it does not start from scratch.
    If the website considers the security benefit \name brings
    outweighs the overhead of extending its firewall, it may still consider \name a viable option.


    \textbf{Performance Overhead:}
    \Name adds the overhead of cryptographic operations at the client side.
    Specifically, all users of an \name-enabled website benefit from
    the passive attack defense at the cost of the encryption/decryption overhead,
    because \name prevents passive attacks without any client-side operations.
    Besides, the overhead of signature verification only affects users who installed the extension and benefit from the active attack defense. The extension does not conduct verification when it detects the user is visiting a website without \name as described in \S~\ref{Design:IntegrityVerifier}.
    Despite the overhead, \name retains the performance benefit of CDNs because it does
    not encrypt public data and CDNs can still cache them at edge servers.
    For private data, \name enables CDNs to cache them in an encryption format within an \name session. The session lives across multiple page visits as described in \S~\ref{Design:EncrytionChannel}.
    Furthermore, the computational overhead can be reduced via optimization techniques at the cost of
    increased design and implementation complexity.
    One technique is to replace a pre-signed digital signature of a web object
    with a hash value and distribute one signed file that includes
    a list of hash values.
    Overall, We believe the current design and implementation of \Name achieves
    a proper trade-off among performance, deployability, and usability, and
    the overhead is acceptable for most websites and users according to our evaluation in \S~\ref{ClientPageLoadTime} and \S~\ref{InternetPageLoadTime}.

    \textbf{Third-party Service Security:} An \name-enabled website (\eg, \url{bank.com}) may use third-party services (\eg, \url{service.com}) but \name cannot protect requests to \url{service.com} if \name is not deployed on the service.
    When a page of \url{bank.com} sends a request to \url{service.com} and the URL is listed as private in \texttt{configure.js} of \url{bank.com}, the client proxy of \url{bank.com} can intercept the request but will forward the request without encryption. It is because \url{service.com} does not distribute a new public key in a specific TLSA record. The request can be protected only when \url{service.com} also deploys \name. We consider such a decision on the third-party service reasonable because a website has no responsibility and ability to protect the data of other services.

    \textbf{Attacks on Service Worker and CSS:} Recent research has explored attacks on Service Worker~\cite{CCS18-PWA,NDSS19-WebPuppet,ACSAC20-SWXSS}. However, these attacks exploit the vulnerabilities in the Javascript code of a website. As discussed in \S~\ref{Assumptions}, such attacks are orthogonal to our work since we assume the website's code is trusted. If an attacker modifies a website's JavaScript code to include a vulnerability, \name can detect it through signature verification.

    Besides Service Worker, existing research shows that CSS can be vulnerable~\cite{CSSExfil,CSSInject}. However, these attacks all exploit specific vulnerabilities in a website's HTML or JavaScript code. Although \name does not sign CSS files by default and CDNs can modify CSS code, we are not aware of how to launch such attacks with CSS injection alone and without vulnerabilities in HTML or Javascript code. Furthermore, we can configure CSS to be signed in \name if future research finds that CSS alone can become an attack vector.
\end{mrlong}

\section{Related Work}\label{RelatedWork}

\textbf{No protection of session keys:} Cloudflare has deployed a
solution called \emph{Keyless SSL}~\cite{CloudflareKeylessSSL,stebila2015analysis} that allows a website to keep its TLS private key
at the cost of involving the web server for every TLS connection.
However, it still reveals the TLS session key to the CDN. In this solution, the CDN forwards TLS handshake messages from
a client to the website that holds the required private key.
Keyless SSL requires the website to host a key server locally
to decrypt or sign the handshake messages
forwarded by a CDN. Thus, a CDN that does not know about the private
key can complete the handshakes.

Akamai also has a similar
patent~\cite{gero2011terminating}, and WASP is a similar approach
proposed by Goh et al.~\cite{modadugu2002design}. Besides Keyless SSL,
a solution proposed by Liang et al. adopts DANE~\cite{RFC7671-DANE} to inform clients about a
website's delegation of a CDN provider so that clients will accept the
CDN's certificate~\cite{liang2014https}. Overall, these solutions do
not protect users' private data since a CDN still obtains the session
key of an HTTPS connection.


\textbf{TEE-based solutions:} Another line of research utilizes
a technology called a Trusted Execution Environment (TEE), for
example, Intel Software Guard Extensions (Intel
SGX)~\cite{anati2013innovative,costan2016intel}.
STYX~\cite{wei2017styx} and
Harpocrates~\cite{ahmed2018harpocrates} are two such solutions based
on Intel SGX, but the same problem with Keyless SSL exists: the CDN
still knows the HTTPS session keys.  Herwig et al. designed and
implemented the first true ``Keyless CDN'' called
\emph{Phoenix}~\cite{Security20-Phoneix} using Intel SGX. Their work
achieves the goal of retaining the full functionality of a CDN while
keeping both a website's TLS private key and an HTTPS session key from
the access of a distrusted CDN. However, deploying Phoenix with its
security guarantees requires that a CDN upgrade all its edge
servers and software to support Intel's SGX.  The deployment cost
may discourage a CDN from migrating to such a solution (\S~\ref{DeploymentCost}).
Besides the financial cost, Phoenix's edge server
throughput is three to four times lower than a server without using
SGX (\S~\ref{ServerPerformance}). This is a potential performance
bottleneck that will reduce a CDN's benefits for accelerating web
access (\S~\ref{ClientPageLoadTime}).
\mr{\Name shows a lower overhead than Phoneix in our evaluation
    but does not retain the WAF functionality of a CDN.}

\textbf{Two-domain solution:} Another solution is to use two
separate domain names for CDN-visible content and private content,
respectively. However, this solution faces a few security,
performance, and usability challenges.  First and foremost, current
websites prefer to use CDNs to deliver their base HTML files for page
load acceleration~\cite{AkamaiHTMLDelivery}, as shown in our
measurement (\S~\ref{Motivation}).  Thus, the two-domain solution
cannot prevent an active attacker in a CDN from tampering with the
base HTML files to expose private content.



Besides the security concern, the two-domain solution prevents a CDN from caching encrypted private content, such as encrypted private photos, because the private content is in unshared TLS sessions and CDNs cannot differentiate between HTTP requests. In contrast, \Name retains the current TLS session practice between websites and CDNs, so CDNs can cache the encrypted HTTP body and headers without access to the content.

In addition, websites have to take a few extra steps to deploy the two-domain solution. They include: (1) A website that already shared its private key to a CDN
needs to revoke the certificate and reapply for two different TLS
certificates. Such a procedure could be expensive and
time-consuming for business certificates~\cite{liang2014https}.
(2) The website needs to negotiate two different types of services
with its CDN provider (one for caching and the other for forwarding~\cite{ProlexicRouted})
(3) The domain separation may require extensive web restructure and
modification to include the new domain in the URLs.

\textbf{CDN-on-Demand:}
Gilad et al. use the two-domain solution to build a low-cost on-demand
CDN~\cite{NDSS16-CDNonDemand}. In their design, they force a client to
fetch all base HTML files from the origin server through the private
domain.  Although this design prevents active attacks, it is
incompatible with the current practice that websites use CDNs to
deliver the base HTML files~\cite{AkamaiHTMLDelivery} and will increase
the page load times of websites' landing pages.  Finally, it inherits
the other drawbacks of the two-domain solution in terms of
functionality and deployment as discussed
above.

\textbf{TLS modification:} Researchers also proposed to modify TLS
to make middleboxes visible in the TLS handshake. Naylor et
al. proposed mcTLS~\cite{SIGCOMM15-mcTLS} to provide different context
keys and use these keys to control what content middlesboxes can read
or write to. Bhargavan et al. discovered the security vulnerability of
mcTLS and provided an alternative to it with formal
proof~\cite{SP18-formal}. Lee et al. extended mcTLS to maTLS,
which makes middleboxes auditable~\cite{NDSS19-maTLS}.  Compared to
\name, such solutions face significant deployment challenges as
they modify the HTTPS/TLS protocol stack.
End users need to explicitly authorize eligible middleboxes' certificates
in the TLS handshake, which may
raise usability concerns.  Clients, servers, and CDNs all need to
upgrade their TLS libraries and adapt their application code to use
the new protocol.



\textbf{Summary:} We consider \name strikes a unique balance among
privacy, performance, user interface, and deployment costs.
\ifx\EXTEND y
    Table~\ref{tab:changes} in the appendix compares the features among the various solutions.
\fi
Compared to mcTLS and the two-domain solution, \name does not change the current web interface and requires fewer changes in the
web ecosystem. Compared to the TEE solutions, \name does not require a
hardware upgrade by a CDN provider.

\section{Conclusion}\label{Conclusion}
We have presented \Name, a system that allows a website to use a CDN
for DDoS protection and web acceleration without exposing the
sensitive data it exchanges with its users.  \Name encrypts sensitive data
transmitted between the client and the origin server so that a distrusted CDN
cannot eavesdrop on their communications.
\Name introduces low overhead, and it is easy to deploy.
A unilateral deployment by a website can prevent a passive eavesdropper in a CDN.
If a user installs \name's browser extension, \name can prevent an active
attacker inside a CDN from eavesdropping on or tampering with their private
communications.

\begin{acks}
    We sincerely thank anonymous reviewers for their detailed comments. We are also grateful to Jeffrey Chase, Matthew Lentz, Bruce Maggs, Kartik Nayak, Michael Reiter, and Danyang Zhuo for their insightful discussion and feedback. This work is supported in part by NSF award CNS-1910867.
\end{acks}


\bibliographystyle{ACM-Reference-Format}
\ifx\EXTEND y
\else
    \balance
\fi
\bibliography{main}

\ifx\EXTEND y
    \appendix
    \appendix
\section*{Appendix}

\begin{table*}[ht]
    \caption{\label{tab:changes}\textbf{\small This table shows the
            changes that each solution introduces to the components of the
            current web ecosystem. The symbol `-' means no change is
            required by a component. $^*$ Chrome/Chromium needs a patch as
            it misses a standard function API.}}
    \begin{center}
        \begin{tabularx}{\textwidth}{c|X|X|X|X}
            \tabhdr{Components} & \tabhdr{\Name}                                               & \tabhdr{TEE~\cite{Security20-Phoneix,CoNEXT17-mbTLS}}        & \tabhdrTwo{TLS modification}{\cite{SIGCOMM15-mcTLS,SP18-formal,NDSS19-maTLS}} & \tabhdr{Two domains}                           \\
            \hline
            Browser             & Extension installation$^*$                                   & \None                                                        & Code modification \newline \& TLS library upgrade                             & \None                                          \\
            \hline
            CDN                 & \None                                                        & Code modification \newline \& TEE-enabled CPU                & Code modification \newline \& TLS library upgrade                             & Negotiate with CDNs for new forwarding service \\
            \hline
            Server              & Reverse proxy configuration                                  & \None                                                        & Code modification  \newline \& TLS library upgrade                            & Restructure website for two domain names       \\
            \hline
            TLS                 & \None                                                        & \None                                                        & Protocol modification                                                         & \None                                          \\
            \hline
            Web pages           & JavaScript library adoption                                  & \None                                                        & HTML modification                                                             & HTML modification                              \\
            \hline
            Security            & Active-attack-resistant \newline \& Passive-attack-resistant & Active-attack-resistant \newline \& Passive-attack-resistant & Active-attack-resistant \newline \& Passive-attack-resistant                  & Passive-attack-resistant
        \end{tabularx}
    \end{center}
\end{table*}

\section{Key Exchange Protocol}\label{Appendix:Handshake}
The protocol uses Diffie-Hellman Key Exchange
(DHKE), and we denote $g$ and $p$ as the system-wide DHKE
parameters. Note that the exponentiation of $g$ is implicitly modulo by $p$.
The protocol works as follows:

\begin{enumerate}
    \item The client proxy generates a nonce $x$ randomly and
          sends $g^x$ to the CDN, which will forward this message to
          the origin server.
    \item After receiving $g^x$, the origin server randomly
          generates a nonce $y$ and computes the session key
          $(g^x)^y=g^{xy}$. As in TLS~1.3, we use an HMAC-Based Key
          Derivation Function (HKDF)~\cite{RFC5869-HKDF} to generate a
          cryptographically strong secret key. The origin server also
          generates a session ID $sID$ and a session expiration time
          $expireT$. The server configures this expiration time and generates a message that includes $g^y, sID,
              expireT$, signs the message, and sends it back to the
          client proxy.
    \item With the response from the server and the public key
          from DNS, the client verifies the signature in the above
          message. After a successful verification, the client proxy
          accepts the common secret derived by HKDF and $g^{xy}$ as the session key.
          Since the public key is only used in this step, the DNS
          query can be sent in parallel with the first key exchange
          message to the origin server.
\end{enumerate}


\section{Implementation}\label{Appendix:Implementation}

We present a prototype implementation of \Name. The
implementation faces the challenges of being compatible with existing
websites and browser implementations. We describe how we address these
challenges to make \name easy to deploy.


\subsection{Client Proxy}\label{Appendix:ClientProxy}



We implement the client proxy library using the \texttt{Service Worker
    API} supported by mainstream browsers, including Chrome, Firefox,
and Safari. A Service Worker is a JavaScript file that runs in the
background of an open web page. A Service Worker enables the client
proxy to intercept each HTTP request that a browser issues. When a
browser initiates an HTTP request, either by Asynchronous JavaScript
(AJAX) or by navigation to another URL, the browser will trigger a
\texttt{fetch} event and call the corresponding handler in the Service
Worker script. The handler code can block or modify the request and
the corresponding response. We implement a handler function to encrypt
private requests and decrypt private responses.

The client proxy stores the session state of \name in browser cache through \texttt{Cache API}. Each time when the handler of \texttt{fetch} event is called, it will retrieve the session state from the cache. If the session state does not exist, the handler starts to handshake with the server through a handshake URL specified by the web developers. If web developers enable pre-handshake, the client proxy will
bind another handler for pre-handshake to the \texttt{active} event. This event is triggered when the Service Worker is activated in the landing page.


A website can embed the client proxy JavaScript library and register
the client proxy's Service Worker script (named \texttt{sw.js} in our
library) in a landing page that a user visits before she
authenticates herself to retrieve private content, \eg,
\url{login.html}. The changes are only a few lines of HTML code. A
website specifies its sensitive URLs and a handshake URL in the
configuration file (\texttt{configure.js}). The configuration file
uses regular expressions to reduce the overhead of configuration and
the size of the configuration file. An example of
\texttt{configure.js} is shown in Figure~\ref{fig:configure.js}.



\subsection{Server Proxy}\label{Appendix:ServerProxy}

We implement the server proxy as a module of a widely used web server:
NGINX~\cite{WebServer}. A website can modify an NGINX configuration
file (\texttt{nginx.conf}) to activate the server proxy. Our module
provides three configuration directives:
\begin{itemize}
    \item \textbf{cloakhello} A website uses this directive to
          specify the server path/URL where the code that constructs
          the session key with a client proxy resides. It should be
          exactly the same handshake path/URL as the one specified in
          the client's configuration file (\texttt{configure.js}).
    \item \textbf{cloakenc} This directive allows the website to
          specify the sensitive URLs related to private content. The
          server proxy will automatically decrypt the requests or
          encrypt the responses of those URLs. This set of URLs should
          be the same as those specified in the client's configuration
          file (\texttt{configure.js}).
    \item \textbf{cloakstate} This directive specifies the size of
          the shared memory used by the NGINX processes to store \name-related
          session state. This directive also provides the server
          module the private key corresponding to the server's
          public key distributed via DNS.
\end{itemize}

A website needs to install our NGINX server module and configure it to
deploy the server proxy. If a website already uses NGINX as its web
server implementation or reverse proxy, it can integrate the \name's
server module into their NGINX instances by recompiling the NGINX
source code. If it does not use NGINX, it can compile our module into
an NGINX server, run it as a reverse proxy in front of its web server,
and configure the NGINX proxy to forward all requests to the web
server. A website also needs to configure the three directives we
describe above to customize the server proxy. We show an example of
\texttt{nginx.conf} in Figure~\ref{fig:nginx.conf} for interested
readers.


\begin{figure}[t!]
    \begin{verbatim}
var config = {
    sensitiveURLs: [
        "/transactions" # String matching
        /^\/profile/, # Regular expression 
    ],
    handshakeURL: "/clientHello"
};
	\end{verbatim}
    \caption{\textbf{\small This figure shows an example of the client proxy configuration file \texttt{configure.js}. A website can specify its sensitive URLs succinctly using regular expressions. In this example, the sensitive URLs are those whose paths are exactly ``/transactions'' and start with ``/profile.'' } }\label{fig:configure.js}
\end{figure}

\begin{figure}[t!]
    \begin{verbatim}
# Allocate 10MB shared memory named "shared".
# The private key is located at /cert/private.pem.
cloakstate shared 10240 /cert/private.pem;

# Encrypt the authentication cookie (named token).
cloakcookie token;

# The path for constructing the secure channel
location /clientHello {
    cloakhello on;
}

# URLs with the path "/transactions" are sensitive.
location = /transactions {
    cloakenc on;
    proxy_pass http://origin:port$uri$is_args$args;
}

# URLs starting with the path "/profile" 
# are sensitive.
location /profile {
    cloakenc on;
    proxy_pass http://origin:port$uri$is_args$args;
}

# Other requests are forwarded to the origin.
location / {
    proxy_pass http://origin:port$uri$is_args$args;
}
	\end{verbatim}
    \caption{\textbf{\small This figure shows an example of \Name's configuration directives in an NGINX configuration file \texttt{nginx.conf}. The directive \texttt{proxy\_pass}, provided by NGINX, is used to forward a request to the origin server\protect\cite{NGINXProxyModule}.}}\label{fig:nginx.conf}
\end{figure}

\subsection{Integrity Verifier}\label{Appendix:IntegrityVerifier}
We implement the integrity verifier as an extension to both Firefox
and to Chromium. A user needs to install this extension to take
advantage of the integrity verifier. However, the Chromium implementation raises a complication. Unlike Firefox's \texttt{webRequest API} for extensions, Chromium does not allow an extension to obtain the body of an HTTP response~\cite{FirefoxWebRequest,ChromiumBugs}.
Therefore, we provide a Chromium patch that adds
an API named \texttt{webRequest.onDataReceived} to the Chromium
extension. Thus, Chromium users need to update their Chromium to
install the extension.

The extension registers a callback function through the API to read
the response body and verifies the signature attached to each of the
public executable web objects. The extension also uses a DoH request
to obtain the public key of the website for the verification.


The extension will alert a user if the verification fails. However, a
website may not deploy \name. In this case, the extension should not
falsely alarm the user. Our implementation uses the existence of a
server's public key record in DNS to differentiate the above two
cases. If a website does not provide the public key in its DNS record,
we consider it has not deployed \name, and the integrity verifier will
skip verification.




\fi

\end{document}